*Article*

# Infrastructural Requirements and Regulatory Challenges of a Sustainable Urban Air Mobility Ecosystem

Arpad Takacs [1,†] and Tamas Haidegger [1,2,*,†,‡]

1. Antal Bejczy Center for Intelligent Robotics, EKIK, Óbuda University, 1034 Budapest, Hungary; arpad.takacs@irob.uni-obuda.hu
2. John von Neumann Faculty of Informatics, Óbuda University, 1034 Budapest, Hungary
* Correspondence: haidegger@irob.uni-obuda.hu
† Authors contributed equally to this work.
‡ Current address: Óbuda University, Becsi ut 96/B, 1034 Budapest, Hungary.

**Abstract:** The United Nations has long put on the discussion agenda the sustainability challenges of urbanization, which have both direct and indirect effects on future regulation strategies. Undoubtedly, most initiatives target better quality of life, improved access to services & goods and environment protection. As commercial aerial urban transportation may become a feasible research goal in the near future, the connection possibilities between cities and regions scale up. It is expected that the growing number of vertical takeoff & landing vehicles used for passenger and goods transportation will change the infrastructure of the cities, and will have a significant effect on the cityscapes as well. In addition to the widely discussed regulatory and safety issues, the introduction of elevated traffic also raises environmental concerns, which influences the existing and required service and control infrastructure, and thus significantly affects sustainability. This paper provides narrated overview of the most common aspects of safety, licensing and regulations for passenger vertical takeoff & landing vehicles, and highlights the most important aspects of infrastructure planning, design and operation, which should be taken into account to maintain and efficiently operate this new way of transportation, leading to a sustainable urban air mobility ecosystem.

## 1. Introduction

During the past decade, Vertical takeoff & landing vehicles (VTOLs)— or *flying cars*, as often referred to in popular literature—gained much attention in the scope of modern passenger transportation systems. Expectations are extremely high towards these innovations, as they form the possible key to sustainable urban mobility ecosystem. Nevertheless, like in the case of many other safety-critical novel technologies, their regulations and legislation are being in their embryonic phase. The technological maturity of VTOLs is well reflected through system demonstrations, however, development and legislation processes show similar challenges to autonomous ground vehicle (a.k.a *self-driving car*) development. The recent speed at which robotic technologies advanced (partially due to the coronavirus pandemic-induced global lock downs) also affected VTOLs [1].

Until the technology reaches its full maturity, it is expected that the infrastructure and the physical space domain will be shared by human-operated, traditional vehicles and (semi)-autonomous systems. This includes any discrepancies, changes or modifications to the operational environment, such as weather, traffic, road works or other marked changes in traffic rules. Vehicles sharing this environment need to be able to simultaneously respond to these changes. Arguably, most of the common traffic or environment changes do not affect VTOLs during point-to-point travel when airborne, except for a small number of physical obstacles and wind conditions. Takeoff & landing, on the other hand, open up new challenges specific to VTOLs [2], where the underlying technology used by commercial aviation offers a baseline that has proven itself in the past decades. These challenges, however, apply to low-altitude flight as well, especially when entering a crowded environment.

Unmanned Aerial Vehicles (UAVs) a.k.a. *drones* provide useful good practices for initiatives that are addressing the regulation challenges of VTOLs, and some of the regulations can already be derived from existing documents. Furthermore, it is expected that there will be a significant regulatory overlap with autonomous ground vehicles, based on the common view that VTOLs (fossil-fueled or electric) will be deployed with a high-level autonomy. The engineering community argues that the maturity and safety of the autonomous ground vehicle technology will have been proven by itself at the time of global deployment of VTOLs, provided that the underlying technology allows it. Regulatory, design and safety issues are also widely discussed in the literature, however, the introduction of elevated traffic also raises environmental concerns, and greatly affects the existing and required service and control infrastructure [3].

In the past years, surveys have also evolved in the United States of America, targeting the competition among the different ride-hailing services and passenger transport, from the commuting point of view. Such surveys compare traditional ground vehicles, self-driving cars and electric air taxi services, and highlight the importance of the perception of the individuals and attitudes regarding the introduction of this new way of transportation. Regulatory and technological aspects were also included in such surveys as objective indicators [4].

Commercial drones have also gained popularity recently, with an increasing number of applications in surveillance services, discovery and delivery [5]. In general, countries are allowed to set their own regulations, which focus on the technical requirements (e.g., altitude of flight or design) and day-to-day usage (insurance, training, licensing or flight zone definition) [6]. On the other hand, takeoff & landing sequence regulations are very sparse for VTOLs, where collision avoidance and path planning may play a significant role in the future, provided that these vehicles are expected to rely on suitable runways for specific maneuvers.

Advanced Air Mobility (AAM) market forecasts project difference consumer options for urban air travel, allowing several trends to emerge from the use cases [7]. These trends can be associated with three distinct uses of VTOLs:

- Package delivery: the door-to-door delivery of goods and food for the consumer market;
- Personal air vehicles: long-range personal vehicles for commuting and recreational use;
- Taxi service: an on-demand urban transportation service as an alternative for ground public transportation.

Other future trends project the rapid adaptation of electric vehicle technology for aerial use cases, developed in parallel with the commercial aerospace industry trends in sustainability, such as electric environmental control, developed recently by Boeing [8]. Promotion of the industry has also gained speed by governmental sector; in the United States, this initiative has already been issued in the Advanced Air Mobility Coordination and Leadership Act. It is expected that regulatory bodies and public entities will raise public awareness of AAM to gain public trust and increase the economic footprint for the VTOL technology.

This paper is intended as a scoping review of the state-of-the-art regulations of urban air mobility, discussing current safety and design issues and infrastructural challenges. In order to highlight the timeliness of the scope of this work and its relevance, the authors have cross-referenced the database of research articles in the largest publication search engines (Google Scholar, IEEE Xplore) with the keywords used. *VTOL* and *VTOL&Infrastructure* keywords provided 17,200 and 3650 matches, respectively; *Urban Air Mobility* is mentioned in more than 4000 papers, while *Autonomous Aerial Vehicle & Safety* are used together 1000 times. The most relevant first few hundred results were analyzed and sampled for this narrated review paper.

## 2. Taxonomy

In recent literature, several terms associated with VTOLs have emerged, and have been defined similarly based on the context of the relevant works. In this review, these terms are used as guidelines, with the following definitions applied to them:

- *VTOL:* Vertical take-off and landing vehicle—an arbitrarily controlled vehicle that can hover, take off and land vertically.
- *Unmanned Aerial Vehicle (UAV)* or *Unmanned Aircraft*: an aircraft in general terms without a human pilot on board. An UAV includes several components of an unmanned aircraft system: a ground-base controller, the vehicle itself and a communication system connecting the two.
- *Drone:* a term interchangeable with UAV, referring to an aircraft without a human pilot operator on board. Traditionally, vehicles not requiring pilots or drivers inside were also associated with this term.
- *Flying car:* a personal air vehicle or an aircraft with road bearing capabilities. The term is commonly associated with means of door-to-door transportation both in the air and on the ground.
- *Autonomous ground vehicle* (for reference:) a ground vehicle capable of L3 (partial) or higher level of automation. In this context, the L3 level refers to the definition of automation levels for self-driving cars by the Society of Automotive Engineers (SAE).

Note that while autonomous VTOLs are fully considered to be robots as per the most recent robot definitions based on standards [9], given their heavily regulated origin domain of aircraft, they are seldom regulated together with general robotic devices.

## 3. Overview of Existing Regulations and Guidelines

On the global level, national bodies are obliged to individually set the regulations. Today, these are formalized along operational aspect in four attributes:

1. Drone mass;
2. Population density;
3. Altitude;
4. Use-case.

Regulations have been formally collected into six distinct approaches based on a nation's attitude toward the commercial use of UAVs. Future regulations targeting VTOLs can rely on these guidelines, however, the components are not applicable to self-driving cars and automated ground transportation [6]. The most common, articulated regulatory approached being:

- Outright ban: Commercial use of drones is not allowed.
- Effective ban: Commercial drone licensing is allowed by existing formal processes. However, it is often impossible to fulfill the requirements and licenses are systematically disapproved.
- Requirement for constant visual line of sight (VLOS): The drone can only be operated within the pilot's visual line of sight. The operational domain is limited, as well as the altitude and range of operation.
- Experimental uses of beyond visual line of sight (BVLOS): a moderately strict VLOS with clearly defined exceptions when the pilot is allowed to operate without direct line of sight. Requirements are set to the pilot ratings and the use case.
- Permissive: commercial drone use is not restricted for most use cases. Guidelines for the operation and licensing requirements are set by a regulatory body. Insurance and registration policies also exist. The drone operation for recreational or commercial use is straightforward if the required procedures are followed.
- Wait-and-see: a strategy that relies on the regulations of other countries, constantly monitoring the outcome. Almost no active drone-related legislation exists in these cases, which often leads to inconsistencies.

The European Union Aviation Safety Agency (EASA) addressed most operation types and assosicate risk levels in the EU Regulation 2019/947. The regulation was applied on 30 December 2020, and defines 3 distinct operation categories [10]:

- Open: low-risk operations, where the drone operator has to comply with the requirements relevant for the operation intended. This ensures safety, and due to the low risk level of this category, there are no requirements for authorization pre-flight. Further subcategories were added to the open operation, though their conceptualization is the same as above.
- Specific: the drone operator has to ensure safety by obtaining authorization for operation. The authorization is issued by a (national) competent authority prior any flight-specific operations. Such operations are considered riskier, requiring the drone pilot or operator to participate in a risk assessment process. Authorization can only be obtained upon successful assessment and liability agreements.
- Certified: safety is ensured by a strict process of the licensing and optional training of the remote pilot. Operations pose a high safety risk. Therefore, authorization requires the certification of the aircraft and the operator as well.

The European Union member states are required to comply with these regulations. Completion of the existing rules and regulation is possible, provided that the original regulatory rules are not softened by addendum. There is no need to re-list the EU-level regulations in the state-level documents, honoring that these documents complement each other. In 2021, EASA issued a set of dedicated technical specifications specifically for VTOLs, collected in a *Means of Compliance with the Special Condition VTOL* document. The Special Condition was subject of a public consultation, and its purpose was to establish safety and design objectives, such as hidden failures, control margin awareness or even the installation of recorders [11]. An example for lift/thrust unit cascading failure evaluation is shown in Figure 1.

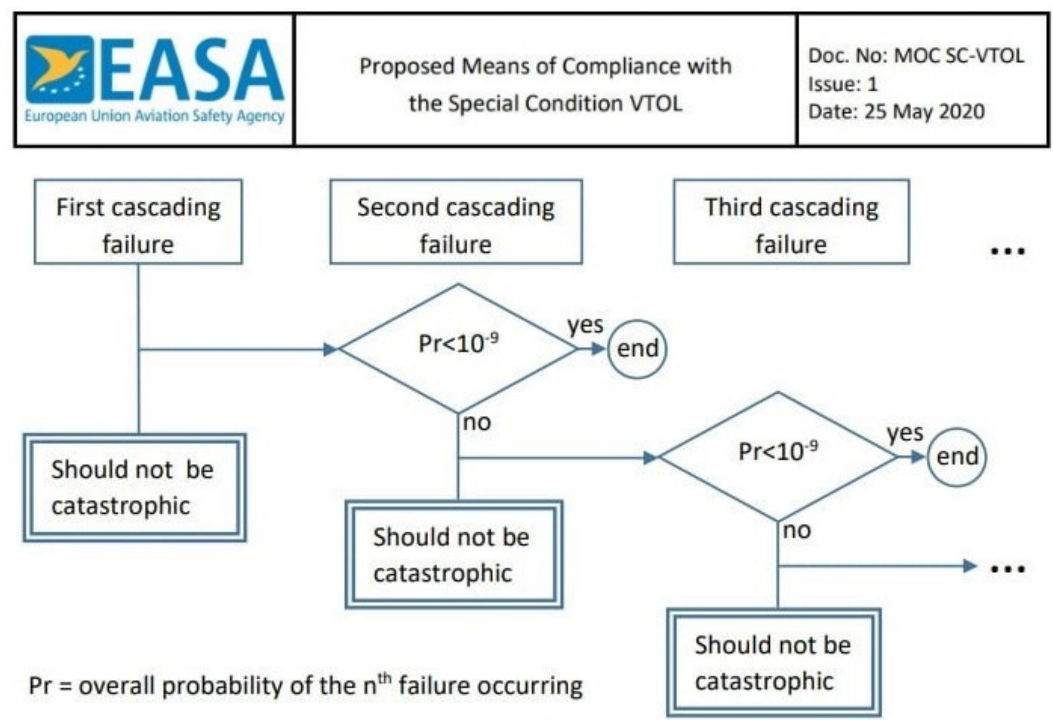

**Figure 1.** Methodology for the lift/thrust unit cascading failure evaluation. *Image credit: EASA.*

The US Federal Aviation Administration (FAA) announced a set of rules for general Unmanned Aircraft use in December 2020. Operators are allowed to fly over people and at night under special conditions according to the new rules, provided that they possess a Remote Identification (ID) for their drones. At the time of the announcement, more than 203,000 FAA-certificated remote pilots were registered by the system, operating over 1.7 million drones. The data was shared publicly by FAA's year-end press release [12].

## 4. VTOL Prototypes and Development History

In recent years, a growing number of functional flying car and VTOL prototypes have emerged, designed either by industrial or research communities. Furthermore, prototypes are often accompanied by service portfolios, including infrastructure, Transportation-as-a-Service and commercialized delivery. This supports the timeliness and actuality of the review of some of the most relevant initiatives listed below.

The first flying car prototype that was designed to vertical takeoff & landing operations was introduced in 2003. The M400 Skycar was planned for hovering and changing its motion state to rapid forward flight [13]. Direct propulsion to the thrusting fans was provided by Rotapower engines, based on a then-popular Wankel design—the same engine design that was used in Mazda RX-7 sports cars until the early 2000's. The system development stretched over decades, however, no successful demonstration of its *standalone* flight capabilities was presented. Over the past decades, the physics concept and the engineering design accuracy were approved and validated by several companies working on prototypes.

Japanese public–private initiatives gave birth to a new industry branch for electric VTOL (eVTOL) development in 2019, aided by infrastructure and service concepts as well. The most remarkable prototype of this era was created by SkyDrive Inc. (Tokyo, Japan), a start-up company, introducing their third generation SD-03 single-seat eVTOL. The vehicle, which weights 400 kg, resembles a hydroplane, and was presented in the scope of a press conference recently. The size of the prototype is comparable to the size of an everyday-use ground vehicle, though significantly wider. In the mechanical design, material choice for durability and density, as well as the balancing of the structure itself, played an important role [14]. The manufacturer planned several designs to address lighting and visibility considerations, as VTOL structural design principles are still loosely regulated worldwide.

The Dutch company, PAL-V (Raamsdonksveer, NL) introduced a prototype beyond concept named Personal Air and Land Vehicle Liberty (PAL-V Libery), which successfully demonstrated land and air cruising capabilities, as well as safe take-off and landing maneuvers. The first commercially distributed models are already available for pre-order, provided that the buyers possess a nationally approved private pilot license (PPL). The first deliveries are expected in 2022, and the company offers accredited flight instructor support along a purchase [15].

Technology giants and ride sharing companies have also expressed their interests in joining the eVTOL development and operation business, investing in infrastructure and service development. Uber issued a ridesharing program for eVTOLs, setting strict requirements for the qualification of these vehicles. 8 OEM partners participated in Uber's program, designing prototypes that adhere to the requirements set by Uber, i.e., the vehicle cruising speed must reach 150 mph and it should carry up to 5 passengers [16]. Uber Elevate was later acquired by Joby Aviation and its partner network was extended in December 2020, by the involvement of the parent companies. Uber also launched partnerships with the FAA and NASA (National Aeronautics and Space Administration) federal/government agencies in the United States, targeting eVTOL fleet operations and the corresponding air traffic management systems. Real estate development companies Oaktree and Hillwood, with engineering communities openly supported the initiative in order to establish *skyports*. Skyports are intended for loading and unloading passengers; requirements for recharging capabilities, capacity limits and space uptake were also issued by Uber on pair with the vehicle design requirements. As of January 2022, Joby Aviation's second aircraft received an FAA Special Airworthiness Certificate and US Air Force airworthiness approval, aiming for commercial operation kick-off in 2024 [17].

While eVTOLs are popular among smaller, lightweight prototypes, gasoline-powered drones have also been introduced lately by some traditional manufacturers (Figures 2 and 3). The Kawasaki K-Racer X1 was first tested late 2021, intended to carry payloads up to 100 kg. Experts argue that the fossil-fueled VTOLs have significantly larger range capabilities compared to their electric counterparts. Kawasaki is also working on larger rolling modules to integrate with factories and delivery centers, but the size of the cargo robot may allow manned flights as well in the near future [18].

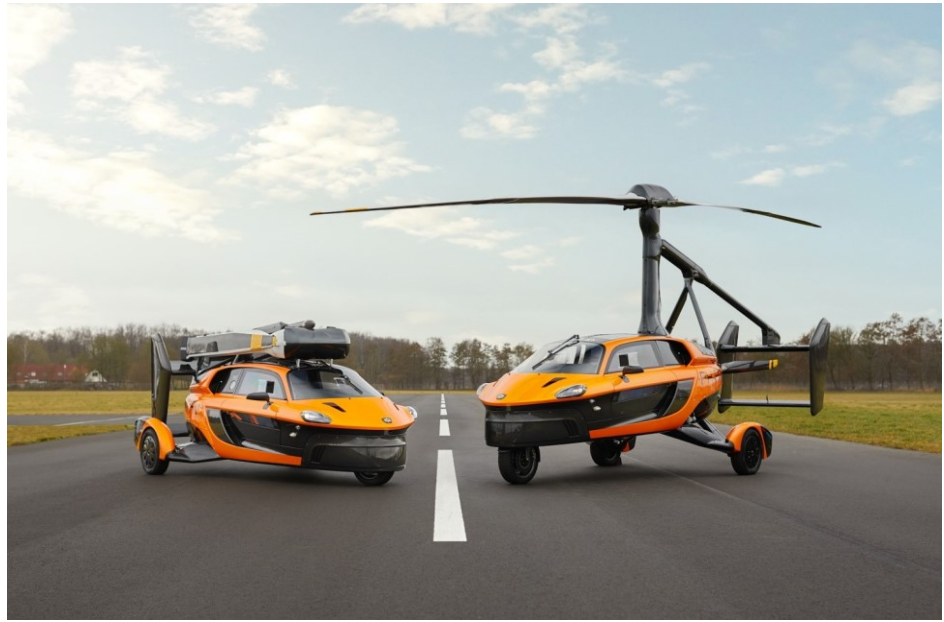

**Figure 2.** The Pal-V Liberty flying car in both ground use (left) and flying (right) setups. *Image credit: PAL-V*.

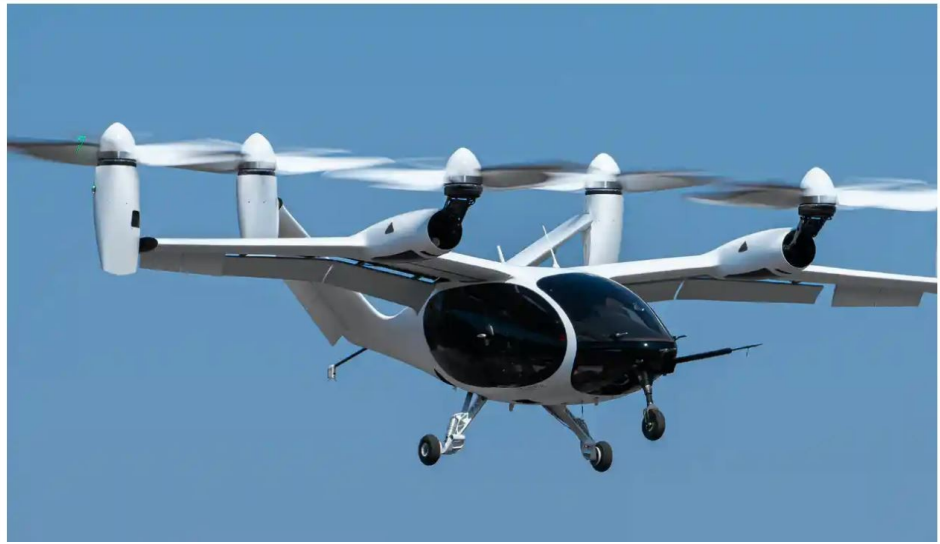

**Figure 3.** Uber Elevate's concept eVTOL, intended for operation under the joint complete transportation service with Joby Aviation. *Image credit: Uber Air*.

## 5. Safe and Sustainable Development

Commercial VTOL applications need the support of unified regulations and policies. The demand for such policies is strong among developers and future operators, because without such guidelines, sustainable development and continuous improvement may suffer. The maturity of the VTOL technology has reached a level where engineering design evolves hand-in-hand with regulations. The first commercially available VTOLs are expected to be semi-automated, and their worldwide availability is expected at the end of 2030 [19]. Ground vehicle automation has increased the acceptance and public awareness of the underlying technology, which, from the viewpoint of VTOL development, may serve as a roadmap for the next stages of technology readiness level. Expectantly, regulatory administration will be accelerated by the simultaneous development of novel air and ground transportation modes. On the other hand, the technology acceptance is subject to the perception received regarding adverse safety incidents. Media communication and the education of the public about the limitations and capabilities of novel technologies may reduce the negative impact of failure events, e.g., recent accidents by Tesla or Uber [20].

With the growing number of VTOLs in the traffic infrastructure, the regulation of VTOL landing and takeoff is gaining much attention from the safety and infrastructure design point of view. Risk analyses need to be conducted to fulfill the requirements of each specific national airspace agency [21]. Such agencies and systems (e.g., the National Airspace System—NAS in the US) oversee airspace, navigation, services, rules and policies for air traffic, and their mission is to protect ground installations and establish a safe airspace environment for all civil, military and commercial aviation. Naturally, VTOL operation and design are often approached from the traditional aviation regulatory view. The focus is on system level redundancy, ensured by the quality requirements for all subsystems, such as motion control or routing. In general, physically separated, varying principle backup systems ensure this redundancy. Moving to a safe state due to malfunction or environmental effects is a necessary feature for all VTOLs, manned or autonomous, and the requirements for such maneuvers can be directly derived from the autonomous ground vehicle SAE L4 highly automated systems [22]. Private air traffic controllers are also invited to mandate standards in specific countries and states, provided that the minimum safety standards have already been issued by the primary regulatory agencies and bodies [23].

Regulatory challenges and directions are mainly derived from the technological readiness and the capabilities of the infrastructure. In terms of safety, VTOLs are heavily affected by the weather conditions of the operation domain. In the case of ground vehicles, adverse conditions cause degradation in vision and other sensing modalities of both the human pilots and autonomous systems, which has been addressed in detail in the past decade by providing models for information degradation [24]. In addition to these disturbances, VTOLs are also subject to the loss of motion stability while airborne. Industry standards in both aviation and automotive industries, hardware-in-the-loop (HiL) testing is aided by certified simulation tools, where borderline cases of the operation domains can be investigated. Systematically setting parameters for precipitation type and intensity, wind attributes and other conditions of visibility ensures a throughout test process for the issues mentioned above. These environments rely on advanced environment models and account for the detailed dynamics and motion model of the tested vehicles, which makes them an excellent choice for pilot training and certification in case of manual or semi-autonomous operation. This inspired the formulation of test centers focusing on certification and training as well, where the pilots learn a unified software for monitoring and control of drones in virtual environments (https://mydronespace.hu/, accessed on 2 April 2022).

*SAE Levels of Autonomous Vehicles*

The current technological maturity of autonomous vehicles, both ground and airborne, allows developer communities and regulatory bodies to initiate discussion on the level of autonomy of these systems. Categorizations allow a long-term planning for both the legislation and the supporting infrastructure, and often derived from the intended use case regarding the roles of human operator and the autonomous system in terms of decision making, motion control and comprehensive environment perception. In the US, the first such categorization was created by the National Highway Traffic Safety Administration (NHSTA), published in 2013 [25]. The definition was later extended and revised by the Society of Automotive Engineers (SAE). SAE recommended 6 autonomy levels in the J3016_201609 taxonomy, focusing on coherence and consistency [26,27], being applicable to a wide range of autonomous systems.

- *Level 0: No driving automation.*
  The vehicle is not capable of performing any automated actuation in any of the use cases. Warning signals may still be transmitted to the human operator.
- *Level 1: Driver assistance.*
  Advanced driver-assistance (ADAS) is allowed through the automated operation of longitudinal or lateral motion control. The human operator is actively engaged.
- *Level 2: Partial driving automation.*
  The vehicle is fully actuated by the autonomous system in a limited operation, allowing for the joint operation of longitudinal and lateral control. The human operator is actively engaged, so that control takeover can be managed immediately.
- *Level 3: Conditional automation.*
  The human operator is actively supervising the automated system, which is allowed control and end-to-end dynamic driving task. The system is equipped with environment recognition, decision making and control capabilities.
- *Level 4: High automation.*
  Control, decision making and environment recognition tasks are jointly performed in the dynamic driving task in a bounded operational design domain (ODD). The vehicle is capable of moving to safe state in the case of malfunction or too complex driving tasks.
- *Level 5: Full automation.*
  All dynamic driving tasks in all operational design domains are carried out by the automated system without the need of human intervention.

In parallel domains, e.g., the likewise safety-critical field of medical robotics, similar categorizations have been introduced successfully [28]. The 6-level automation levels for drones was first introduced by Drone Industry Insights, relying on SAE's taxonomy as a baseline [29]. The authors listed examples from real life (unmanned) drone applications, and suggested the categories identified in Figure 4. According to the experts common view, the main limiting factors to achieve the more complex levels of the automation of drones are the hardware boundaries of the equipped processing units and the data transfer bandwidth and latency. In addition, LoA 5 authorization is still missing from existing regulation of commercial UAVs.

- *Level 0: No drone automation.*
  The UAV is operated by the human pilot at all times, all control aspects and movement are supervised by the human operator.
- *Level 1: Pilot assistance.*
  Due to safety reasons, the overall control and monitoring of the system are done by the human operator. One key function may be delegated to the automated system.
- *Level 2: Partial automation.*
  The UAV is allowed—under the human pilot's supervision—to autonomously control its speed, altitude or heading. Still the human pilot is responsible for the operation safety.
- *Level 3: Conditional automation.*
  The human operator needs to be ready for and immediate takeover, however, the UAV is capable of environment perception and motion planning.
- *Level 4: High automation.*
  The UAV is equipped with a redundant failure-proof backup system, which can take over and move the drone to a safe state. All control and navigation functions are automated.
- *Level 5: Full automation.*
  All aspects of the flight cycle are managed by the automated system, including lift-off and landing, flight maneuvering and operative mission, such as mapping, inspection, repair or delivery.

THE 5 LEVELS OF DRONE AUTONOMY

| Autonomy Level | Level 0 | Level 1 | Level 2 | Level 3 | Level 4 | Level 5 |
|---|---|---|---|---|---|---|
| Human Involvement | | | | | | |
| Machine Involvement | | | | | | |
| Degree of Automation | No Automation | Low Automation | Partial Automation | Conditional Automation | High Automation | Full Automation |
| Description | Drone control is 100% manual. | Pilot remains in control. Drone has control of at least one vital function. | Pilot remains responsible for safe operation. Drone can take over heading, altitude under certain conditions. | Pilot acts as fall-back system. Drone can perform all functions 'given certain conditions'. | Pilot is out of the loop. Drone has backup systems so that if one fails, the platform will still be operational. | Drones will be able to use AI tools to plan their flights as autonomous learning systems. |
| Obstacle Avoidance | NONE | SENSE & ALERT | SENSE & ALERT | SENSE & AVOID | SENSE & NAVIGATE | SENSE & NAVIGATE |

**Figure 4.** UAV regulations: a similar classification to the one defined by SAE for surface vehicles. *Image credit: Drone Industry Insights*.

Traditionally, UAVs were considered compact, agile systems. On the contrary, VTOLs are subject to regulations in the infrastructure, aimed to carry out tasks with high complexity. The definition of the autonomy levels needs to be extended to the takeoff & landing aspect, as well as to transition to a safe state in case of partial system failure or human handover. Cruising, height control and collision avoidance are few additional topics to be addressed, especially at the middle autonomy levels, where joint operation is allowed for the automated system and the human pilot too. The gradual introduction of VTOLs into the infrastructure ensures safety and sustainability for the increase of VTOL use in modern transportation, where the following guidelines for VTOL autonomy levels are proposed (Figure 5).

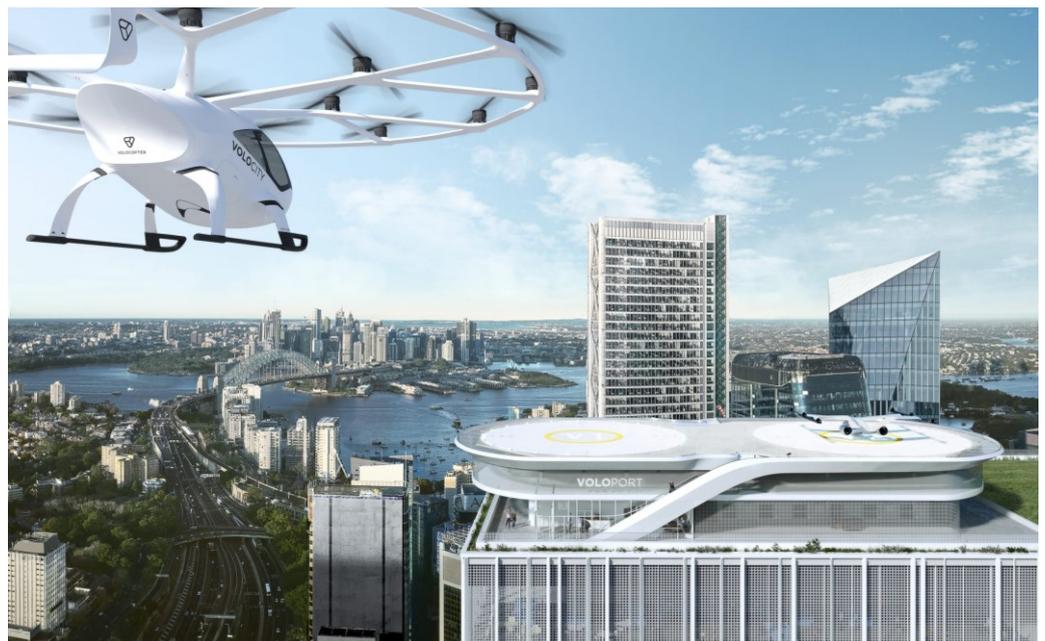

**Figure 5.** VoloPort vertiport concept by Volocopter, designed for urban eVTOL operations. *Image credit: Volocopter*.

- *Level 0: No VTOL automation.*
  The human pilot is controlling the heading, speed and altitude. The vehicle may be equipped with semi-autonomous motion stabilization system. During *levitation*, the VTOL is maintaining its absolute position and orientation. However, it is subject to noise generated by load distribution or wind. In order to maintain safe operation at all times, minimum requirements should be prescribed in safety regulations for such non-autonomous VTOL design specifically.
- *Level 1: Pilot assistance.*
  An automated approach may be applied to VTOLs for critical maneuvers, with the addition of cruising, and expecting that autonomous control in general will be present in VTOLs in Level 1 and above for safety reasons. The independently controlled parameters—heading, altitude and cruising speed—can be further adjusted by the human pilot, provided that the system may support it at any time.
- *Level 2: Partial automation.*
  The human operator supervises the automated takeoff & landing sequence, which is now executed fully by the autonomous system. Joint operation is possible in case of emergency. Autonomous cruising is implemented on a predefined trajectory set by key-points or coordinates. Visual path planning is not applicable at this level (on the contrary to ground autonomous vehicles). However, collision avoidance with static or dynamic objects is implemented, requiring human supervision at all times.
- *Level 3: Conditional automation.*
  The human pilot does not supervise the takeoff–cruising–landing sequence. In addition to Level 2, the VTOL automated system is capable of online trajectory modifications in case of collision avoidance or increased efficacy. The pilot is required to take over control if the VTOL leaves its predefined operational domain. The handover request should be initiated by the VTOL in advance so that the minimum level of situation awareness can be reached at the time of the handover [22].
- *Level 4: High automation.*
  In a well-defined operational design domain, the VTOL may carry out all movement tasks autonomously without the need of human pilot supervision. Complex decision making, emergency maneuvering and unexpected event handling are all built-in, the human operator is not required to hold a license for operation. Should the circumstances force the autonomous system to leave its operation design domain, it may initiate a landing sequence in order to move to a safe or stable state, which includes levitation. Upon moving to the ground, the vehicle may proceed in a ground vehicle mode, if the environmental circumstances allow.
- *Level 5: Full automation.*
  The VTOL is not equipped with interfaces for human control neither for air nor for ground operations. A hybrid air-ground operation is possible and optimized by the autonomous system itself. Multiple landing and takeoff sequences can be executed by the vehicle.

## 6. Impact on Buildings, Infrastructure and the Environment

In 2019, German aircraft manufacturer Volocopter unveiled an air taxi airport concept, aided by an ecosystem of air taxi service and VoloCity aircraft operation. The transportation, as a service, piloted in Singapore and is expected to become operational in the next few years. As of January 2021, FAA accepted Volocopter's application for the EASA certification, which leads to feasibility studies for launching Volocity in several major cities of the United States [30].

In terms of navigation and optimal path planning, VTOLs have a clear advantage in efficiency compared to ground passenger and freight vehicles, significantly decreases the time of journeys. Traffic, topography and geography, along with the restricted road geometry heavily affect traditional ground transportation lines. These constraints do not bind the operation of VTOLs, while in average 60% of the travel distances can be saved by using direct fly paths. The key elements of the aerial journeys are the takeoff & landing: the infrastructure allowing these episodes of flight plays a crucial part in the increased efficiency. In addition, the parking and storage of the VTOLs have to be further solved.

In current taxonomy, takeoff & landing and VTOL storage facilities are referred to as *vertiports*. Regulatory bodies are actively discussing the processes of certification and standardization of these sites, often deriving rules from enacted, already existing guidelines and standards [31]. takeoff & landing lots for commercial and service helicopters, called high-rise *helipads* and conventional airstrips and landing sites serve as examples and starting points for these guidelines. Designing vertiport layouts and their specifications requires optimization and extensive urban planning, which often requires simulation techniques and heuristics. The cost functions of optimization highlight passenger safety and efficiency, often aided by mathematical tools such as Monte Carlo simulation [32].

Buildings that are part of the VTOL ecosystem infrastructure need to be equipped or extended with special facilities and support units. Due to the new way of transportation within the urban area, purpose-built rescue facilities and emergency units need to be established. The storage, in-building transportation of VTOLs, as well as maintenance workshops and charging pads also need to be accounted for. Some of the design considerations may be taken over from existing heliport design guides, however, in a use-case driven development, new terms and local regulations may apply to vertiports.

When urban zones reach a critical number of active airborne VTOLs, the complexity of traffic control significantly increases, especially in case of point-to-point direct flight paths. In order to optimize traffic flow, reduce noise pollution and mitigate damage, a proposal from NASA suggests flight corridors to be located over low-risk areas, constrained by coordinates [33]. Consequently, eVTOL manufacturers are required to adhere to strict noise standards, which reportedly poses one of the biggest challenges for eVTOL design, especially for vehicles relying on the tilted rotor operation principle. The Joby Aviation designs, however, use anhedral tip blades, which significantly improves hovering performance and reduces noise [34].

The cost efficiency and sustainability factors must match new boundary conditions for the trajectory planning problem in 3D, especially when taking the functional range of flight into consideration. The availability of landing and takeoff sites, the geometry and dimensions of the VTOL, along with its aerodynamic properties extend the list of requirements, which are dominantly addressed through simulation and testing. The operation strategy is also affected by the efficiency of the flight path planning at the service providers, such as Lyft or Uber. These companies, undoubtedly, are active participants of shaping concepts and advising regulators, as their ground ride-sharing and ride-hailing services can be extended to the air in the next step of technology evolution. However, ride-sharing services may have their dedicated airspace in the future due to their distinguished mode of operation, the forming regulations prefer an integrated, holistic approach for the usage of the metropolitan airspace by VTOLs, based in an equal sharing and general regulations [35].

eVTOL fuel cell and battery charging introduce another constraint for the policies and regulations related to the infrastructure. Battery life cycle, cost efficiency and energy density have played a key role in the electric ground vehicle development in the past two decades, and the role of these factors reflects similarities in the eVTOL industry as well. Nonetheless, the market size is estimated $498 m in 2019, with a forecast of $3,583 m by 2030, based on the anticipated interest of stakeholders, which indicates a near-future commercial introduction from the ROI point of view [36]. From technology aspect, (e)VTOLs are dominantly designed with split power engines, which allows separation of the rotation speed and force [37]. The new policies need to address the maneuverability and navigation properties of the vehicles, which requires the synchronization of both mechanical and infrastructure design and adhering to the technical requirements.

All VTOLs under Level 3 of automation require an easy-to-operate Human–Machine Interface (HMI), both from the aspects of control and environment perception. This might carry resemblance of the existing ground transportation vehicles, suggesting a hybrid

ground–air use. In such hybrid mode, vehicle operation needs to account for ground obstacles that need to be avoided by flight [38], whether they may be static or dynamic [39]. The human pilot needs information about additional markings, topology, artifacts and corridors. Creating and maintaining these artifacts, such as *levitating* road-side markers, is not an efficient and unambiguous solution, which highlights the importance of the introduction of augmented reality based HMI and advanced vehicle to anything (*V2X*) communication protocols. Remote operation of VTOLs may also benefit from such solutions, for example, in the case of emergency, when a ground control unit takes over the control and moves the vehicle manually to a safe state. Head-up display-based navigation, air–ground communication protocols and the underlying software also play an important role in completing regulations for safe and sustainable VTOL development and introduction to urban mobility.

Sustainability has many aspects when VTOLs are considered, which can be derived from the 3 main pillars of corporate sustainability [40]. These are:

- Environmental sustainability—includes the carbon footprint of the production, operation and maintenance of the VTOL vehicles and infrastructure.
- Social sustainability—the approval of VTOL users and operators needs to be maintained high, including user and/or customer support, fair treatment of employees and social consciousness, trust in technology and the monitoring of the high social impact of the activities of the supply chain.
- Economic sustainability—the profitability of the VTOL industry, reflected to each of the future segments (delivery, taxi service or personal use), public relations with the governing and regulatory bodies, growth strategy and technological roadmap.

Each of these pillars pose various challenges to this newly emerging industry. However, autonomous and electric ground vehicles, commercial aviation and modern public transportation trends provide a solid ground for best-practices in sustainability for VTOLs and the supporting infrastructure.

Among the many advantages VTOLs are expected to bring to urban mobility, several technological and application challenges still remain. VTOLs have limitations regarding fuel efficiency, though eVTOLs are considered to emit less greenhouse gases than internal combustion vehicles, electric ground vehicles still offer a significantly better efficiency. Regarding the speed of operations, congested urban areas may largely limit the airspeed allowed for such vehicles, making the full-length of the journey less time-efficient, landing and take-off maneuvers included. In addition, due to both path planning and technological boundaries, the low operational altitudes may limit the use of the full spectrum of vertical occupancy of urban aerospace, posing another infrastructural challenge to city planning and environmental sustainability [41].

## 7. Efficient Training for Increased Safety

Pilot registration and licensing, the certification of aircraft and the processes in air traffic control mechanism are governed and managed by risk controls around the world. In the US, the mandate of airborne entity regulations falls within the scope of the US Federal Aviation Administration. It is expected that the capacity of today's traffic control systems would be overloaded at the steep ramp-up of commercially used VTOL operation. Reliable and sustainable operation can only be ensured by the introduction of new functions and processes. Systematic and controlled amendments are needed to the regulations in human- or hybrid operated VTOLs, which require a different approach to the rapidly growing and yet loosely-controlled regulation of smaller commercial and personal drones [42].

Similarly to the high variety of driver's licenses for ground vehicles (cars, trucks, buses or distinguished vehicles), VTOLs introduce a new dimension on the skills required from human drivers/pilots for operation. Different licenses refer to the different, vaguely overlapping set of skills required for driving/flying, including the mode of operation: freight vs. passenger, commercial vs. personal, public roads vs. special environments. With the introduction of VTOLs, additional, combined licenses may be required along with training and regulations, largely affected by the mechanical design and mode of operation. Certificates may be required for airplane-like design, gyrocopter-based VTOLs (such as the PAL-V) or fixed-wing design (Aurora PAV or Dufour Aerospace, https://www.dufour.aero/, accessed on 28 March 2022). On the long-term, these factors may affect the airworthiness and certification processes of the VTOLs, and may restrict their use to specific operation domains based on the license of the human pilots and the level of autonomy [21].

The first introduction of VTOLs to urban transportation is expected to be restricted to extensive urban areas, where both the air and ground traffic infrastructures are highly developed. This operational design domain requires numerous landing and takeoff sites, where landing safety is another crucial aspect to consider in terms of technology and regulations. Assuming adequate positioning accuracy and a unified set of requirements for perception, existing methods offer numerous solutions for automated landing maneuvers [43]. It is expected that the first VTOLs will be partially or fully human-operated, and parts or *episodes* of the operation cycle will be restricted to autonomous operation. Depending on the episode length and nature, there are concepts targeting remote flying and operation of the vehicles from ground control centers, including passive supervision, flight data monitoring, mission control and emergency takeover/instructions [44]. This indicates that not only the design, the autonomous system and the human operator, but the supporting ground control units will require a certification and licensing as well, along with the takeoff & landing sites and the supervision of air and ground traffic. While established and accepted policies for these regulations do not exist today globally or on a national level, it is expected that the growing number of initiatives for the ground–air-ground transition and the process supervision will form a sustainable ecosystem for urban VTOL operation.

## 8. Discussion

Vertical takeoff & landing vehicles are gradually forming the future of urban mobility, similarly to the even-paced expansion of ground electric and/or autonomous vehicles. As the technology evolves, the efficiency, level of automation and public awareness and acceptance bring the date of their commercial availability closer day by day. The future of personal and commercial air transportation in short-range is largely influenced by the development of VTOLs, whereas regulatory and technology boundaries of their wide introduction still haven't reached their maturity. In this review, we addressed relevant safety and regulatory state-of-the-art challenges, technology and design aspects from the viewpoint of sustainability, efficiency, focusing on the most relevant viewpoints to be considered by regulatory bodies and urban planning experts in the coming iterations of the introduction of VTOLs to urban mobility. Taxonomies and the current status of automation safety and licensing approaches were discussed, drawing the conclusion that the currently available guidelines and regulations are still being formed, with a strong intention of reusing best practices from both ground transportation and aviation industries.

Historically, policies and regulations are being formed along the industrial requirements and user experience, and service providers and shipping companies are actively engaging in related discussions, justifying the timeliness of the regulatory discussions and policy issues. This includes, but not limited to a standardized definition of the levels of autonomy for VTOLs, pilot licensing, aircraft design, infrastructure, operation and service. The education of the public about the technology and raising awareness of the capabilities and responsibilities of users and operators play major roles in this process: a path that the society needs to take, learning from the impact and introduction of the autonomous ground vehicle technology.

**Author Contributions:** Conceptualization, A.T. and T.H.; Investigation, A.T. and T.H.; Methodology, T.H.; Resources, T.H.; Writing—original draft, A.T. and T.H.; Writing—review & editing, T.H. All authors have read and agreed to the published version of the manuscript.

**Funding:** The research presented in this paper was carried out as part of the EFOP-3.6.2-16-2017-00016 project in the framework of the New Széchenyi Plan. T. Haidegger's work is partially supported by the ELKH SZTAKI-–OE Cyber-Medical System Development project. T. Haidegger is supported through the New National Excellence Program of the Ministry of Human Capacities. T. Haidegger is a Bolyai Fellow of the Hungarian Academy of Sciences. The work presented forms part of the Sustainable Robotics Initiative (https://www.sustainablerobotics.org/, accessed on 25 March 2022).

**Conflicts of Interest:** The authors declare no conflict of interest.

## Abbreviations

The following abbreviations are used in this manuscript:

| | |
|---|---|
| VTOL | Vertical Take-off and Landing Vehicle |
| SAE | Society of Automotive Engineers |
| FAA | Federal Aviation Administration |
| EASA | European Union Aviation Safety Agency |
| ODD | Operational Design Domain |
| UAV | Unmanned Aerial Vehicle |
| HMI | Human-Machine Interface |
| TaaS | Transportation-as-a-Service |
| NASA | National Aeronautics and Space Administration |
| NHSTA | National Highway Traffic Safety Administration |
| NAS | National Aerospace System |
| VLOS | Visual Line of Sight |